\documentclass[conference]{IEEEtran}
\IEEEoverridecommandlockouts
\usepackage{amsmath,amssymb,amsfonts}
\usepackage{algorithmic}
\usepackage{graphicx}
\usepackage{textcomp}
\usepackage{textpos}
\usepackage{xcolor}
\usepackage{url, hyperref, microtype, subcaption}
\hypersetup{
    colorlinks,
    linkcolor=black,
    citecolor=black,
    urlcolor=black
}
\usepackage{booktabs}
\usepackage[acronym]{glossaries}
\def\BibTeX{{\rm B\kern-.05em{\sc i\kern-.025em b}\kern-.08em
    T\kern-.1667em\lower.7ex\hbox{E}\kern-.125emX}}
\begin{document}

\newacronym{gis}{GIS}{Geographic Information Systems}
\newacronym{ml}{ML}{Machine Learning}
\newacronym{dem}{DEM}{Digital Elevation Model}
\newacronym{dtm}{DTM}{Digital Terrain Model}
\newacronym{dsm}{DSM}{Digital Surface Model}
\newacronym{iot}{IoT}{Internet of Things}
\newacronym{ndvi}{NDVI}{Normalized Difference Vegetation Index}
\newacronym{ndwi}{NDWI}{Normalized Difference Water Index}
\newacronym{gari}{GARI}{Green Atmospherically Resistant Vegetation Index}

\graphicspath{ {./figs/} }

\title{Data Science for Geographic Information Systems\\
    \thanks{This work was financed by the Fundação para a Ciência e Tecnologia (FCT), in the framework of projects UIDB/00066/2020, UIDB/04111/2020, CEECINST/00002/2021/CP2788/CT0001 and CEECINST/00147/2018/CP1498/CT0015 as well as Instituto Lus\'ofono de Investiga\c{c}\~ao e Desenvolvimento (ILIND) under Project COFAC/ILIND/COPELABS/1/2022.}
}

\author{\IEEEauthorblockN{
Afonso Oliveira\IEEEauthorrefmark{1}, Nuno Fachada\IEEEauthorrefmark{1}\IEEEauthorrefmark{2} and
João P. Matos-Carvalho\IEEEauthorrefmark{1}\IEEEauthorrefmark{2}
}
\IEEEauthorblockA{\IEEEauthorrefmark{1}\textit{COPELABS, Universidade Lusófona, Campo Grande 376, 1749-024 Lisbon, Portugal}}
\IEEEauthorblockA{\IEEEauthorrefmark{2}\textit{Center of Technology and Systems (UNINOVA-CTS) and LASI, 2829-516 Caparica, Portugal}
}
\IEEEauthorblockA{Correspondence: afonso.oliveira@ulusofona.pt}
}

\maketitle

\begin{textblock*}{190mm}(-1cm,-6cm)
  \noindent \footnotesize The peer-reviewed version of this paper is
  published in IEEE Xplore at
  \href{https://doi.org/10.1109/YEF-ECE62614.2024.10624902}{\texttt{https://doi.org/10.1109/YEF-ECE62614.2024.10624902}}.
  This version is typeset by the authors and differs only in pagination and
  typographical detail.
\end{textblock*}

\begin{abstract}
The integration of data science into Geographic Information Systems (GIS) has facilitated the evolution of these tools into complete spatial analysis platforms. The adoption of machine learning and big data techniques has equipped these platforms with the capacity to handle larger amounts of increasingly complex data, transcending the limitations of more traditional approaches. This work traces the historical and technical evolution of data science and GIS as fields of study, highlighting the critical points of convergence between domains, and underlining the many sectors that rely on this integration. A GIS application is presented as a case study in the disaster management sector where we utilize aerial data from Tróia, Portugal, to emphasize the process of insight extraction from raw data. We conclude by outlining prospects for future research in integration of these fields in general, and the developed application in particular.
\end{abstract}


\section{Introduction}

In the era of data-driven decision making, the fusion of two prevalent disciplines, Data Science and \gls{gis}, stands at the forefront of transformative innovations. Data Science, with its emphasis on extracting meaningful insights from vast datasets, and \gls{gis}, traditionally employed in mapping and spatial analysis, are converging to create a synergistic relation. This integration is proving to be more than a technological evolution; it is a paradigm shift in our approach to understanding and interpreting spatial information.

In the sections that follow, we will discuss the historical evolution of \gls{gis} and Data Science, dissect the key components defining their intersection, and explore real-world applications that highlight the potential of this integration. This paper aims to emphasize the evolving role that Spatial Data Science plays in reshaping our understanding of spatial decision-making processes.

This work is divided into six sections as follows: Firstly, a historical record of the development of the fields of data science (Section~\ref{sec:data-science}) and of \glspl{gis} (Section~\ref{sec:gis}). Then, an overview of the convergence of the two fields (Section~\ref{sec:key-components}) followed by an illustration of the various applications in different areas of interest that hinge on this convergence (Section~\ref{sec:applications}). Finally, we proceed with a small case study to illustrate the process of extracting insights from raw data (Section~\ref{sec:case-study}), and conclude with some final remarks on the work and prospects on the importance of this synergy for future work (Section~\ref{sec:conclusions}).

\section{Emergence of Data Science}
\label{sec:data-science}

Data science emerged as an interdisciplinary field during the second half of the 20th century in response to the escalating volume and complexity of data in the rapidly advancing digital age. It evolved as a convergence of computer science and statistics, applying statistical techniques and using computational tools to extract valuable insights and knowledge from datasets.

The evolution of data science can be traced through transformative milestones. In its early stages, as machine learning gained traction \cite{flach2012machine, borges1989analysis} and more sophisticated statistical methods were developed \cite{kotz2012breakthroughs, freedman2009statistical}, relational databases and data warehousing technologies became crucial for managing and organizing data \cite{chaudhuri1997data}.

With the concept of data warehousing gaining prominence, emphasizing the need for centralized repositories to store integrated historical data \cite{gardner1998building, kimball1996data}, advances in data mining \cite{fayyad1996data} leveraged these emerging technologies as the backbone for storing and managing larger datasets \cite{chaudhuri1997data}.

Eventually, the emergence of big data \cite{mayer2013big}, accompanied by the widespread adoption of open-source tool ecosystems, such as R and Python \cite{ihaka1996r, mckinney2012python}, reached a critical point. Later data science would evolve into a recognized, specialized, and interdisciplinary field \cite{dhar2013data}, witnessing the big resurgence of deep learning and its integration across industries \cite{goodfellow2016deep}, significantly impacting business decision-making \cite{schoenherr2015data, van2016data}.

Recently, there has been an increasing focus on ethical considerations, responsible AI practices, and fairness in data science applications \cite{diakopoulos2016accountability, mittelstadt2016ethics}. Regardless, data science remains a vital component of decision-making processes in most industries, reflecting the dynamic nature of its continuous development over time.

\section{Evolution of Geographic Information Systems}
\label{sec:gis}

A \gls{gis} is a system designed to capture, store, analyze, manage, and present spatial data. \gls{gis} allows users to visualize, interpret, and understand data in a spatial context \cite{longley2015geographic, burrough2015principles}. Since their inception in the early 1960s, these systems have undergone a notable evolution. Originally conceived simply as computerized tools for land resource management, \gls{gis} has progressed through various stages of development.

During its initial stages, \gls{gis} primarily concentrated on the digitization of maps and the conceptualization of spatial databases. The foundational work carried out by Roger Tomlinson \cite{tomlinson1974geographical} set the stage for the development of the first computerized \gls{gis}. With advances in computer technology, the 1970s and early 1980s witnessed the commercialization of \gls{gis} software, with significant contributions from companies such as the Environmental Systems Research Institute (ESRI).

During the following decades, the incorporation of remote sensing data---satellite imagery in particular---was a pivotal development for enhancing the capabilities of \gls{gis} \cite{merchant2000remote}. Standardization efforts, led by the Open Geospatial Consortium (OGC), aimed to establish interoperability among \gls{gis} platforms. Furthermore, the 1990s also included the first definitions of \gls{gis} as a scientific area of development---GIScience \cite{rocha2019geographic}.

During the early 2000s, \gls{gis} underwent a transition to the web, facilitating greater accessibility to spatial data \cite{chang2008introduction}. Subsequently, the rise of cloud-based \gls{gis} platforms and the popularity of web mapping applications \cite{peterson2005maps}, such as Google Maps, significantly contributed to the widespread recognition of \gls{gis}.

The 2010s and beyond witnessed the advent of mobile \gls{gis} and the growth of location-based services \cite{hussein2011mobile}, harnessing the widespread use of smartphones for real-time data collection and field mapping \cite{khan2012mobile}. During this time, there was also a considerable adoption of big data methods and advanced spatial analytics within \gls{gis} applications \cite{goodchild2016gis}.

\section{Key Components of Data Science in GIS}
\label{sec:key-components}

This section explores and covers key aspects of the synergy between \gls{gis} and data science. It begins with spatial data handling and addresses techniques and challenges in preprocessing and analysis ( Subsection~\ref{sub:spatial-data}). Next, the impact of \gls{ml} in spatial analysis is explored ( Subsection~\ref{sub:machine-learning}), followed by several insights on the use of Big Data technologies in the context of \gls{gis} and the importance of scalable solutions ( Subsection~\ref{sub:big-data}). Finally, the section concludes with the importance of effective data communication in \gls{gis}, offering a brief explanation on spatial data representation and interpretation ( Subsection~\ref{sub:data-interpretation}).

\subsection{Spatial Data Handling}
\label{sub:spatial-data}

Spatial data handling in \gls{gis} involves the application of data science techniques to efficiently manage and process geographical information. Techniques, including spatial indexing, geocoding, and spatial querying, play important roles in optimizing data retrieval and analysis. Spatial Indexing involves organizing data to facilitate efficient spatial queries, optimizing data retrieval processes. Geocoding converts places of interest into geographic coordinates, a fundamental step for mapping and essential for spatial analysis and data integration. Lastly, Spatial Querying allows for the retrieval of information based on spatial relations or conditions, offering valuable insights from spatial data crucial for decision-making processes. This field is deemed so crucial that it is often considered a subset of data science itself---spatial data science \cite{garrard2016geoprocessing, rey2023geographic, comber2020geographical}.

Managing spatial datasets poses significant challenges, encompassing concerns related to data heterogeneity, resolution discrepancies, and the necessity for constant updates. Addressing these challenges involves placing a strong emphasis on data cleaning and preprocessing to uphold accuracy in subsequent spatial analyses. Importantly, these challenges give rise to opportunities for innovation. This includes the development of resilient data integration frameworks and the utilization of \gls{ml} for the automated cleaning  of spatial data, topics that will be further explored in subsequent sections.

\subsection{Machine Learning}
\label{sub:machine-learning}

\gls{ml} has emerged as a game-changing tool within the domain of \gls{gis}, facilitating the extraction of valuable insights from spatial datasets and augmenting traditional spatial analysis \cite{choudhury2023geospatial,casali2022machine}. ML algorithms brought a data-driven and adaptive approach to spatial problem-solving, allowing \gls{gis} professionals to uncover more intricate patterns, make more accurate predictions, and automate once complex and time consuming tasks.

Feature recognition and extraction are \gls{ml} techniques that allow automating the identification of spatial patterns and objects. This capability is particularly evident in applications like image classification \cite{sulemane2022vineyard, lim2019tree}, where ML algorithms discern and categorize features within satellite or aerial imagery, contributing to tasks such as land-use classification and environmental monitoring.

\subsection{Big Data}
\label{sub:big-data}

Traditional \gls{gis} methodologies, designed for smaller datasets, struggle with the scale and variety of data generated nowadays. Large-scale spatial datasets, stemming from sources such as satellite imagery, sensor networks, and social media, require innovative approaches to storage, processing, and analysis.

To address the challenges presented by large spatial datasets, \gls{gis} developers are increasingly embracing Big Data technologies and methodologies \cite{akhund2022analysis, lee2015geospatial, li2016geospatial}. This shift aims to establish scalable and distributed computing frameworks that efficiently process extensive amounts of spatial data. Two critical paradigms in this context are the parallelization of spatial data \cite{zhao2016geographical,werner2019parallel} and the scalability of spatial systems \cite{jhummarwala2014parallel,neuman1994scale,hawick2003distributed}.

Spatial data processing and analysis can be efficiently parallelized across a distributed environment, allowing for the execution of tasks that would be prohibitively time-consuming with traditional approaches, therefore enabling the extraction of more meaningful insights from massive datasets. These distributed solutions can also significantly contribute to the question of scalability, ensuring that \gls{gis} applications can handle expanding datasets without sacrificing performance.

The significance of scalable and efficient data processing in the context of Big Data and \gls{gis} cannot be overstated. As spatial datasets continue to grow, the ability to process information at scale becomes paramount for timely decision-making and insightful analysis. Scalable data processing not only addresses the challenges of handling large datasets more efficiently but also unlocks the potential for real-time \cite{sun2016real,al2003real} and near-real-time \cite{gamba1998gis} spatial analytics, enabling a dynamic response to evolving spatial phenomena previously difficult to achieve \cite{zerger2003impediments}.

\subsection{Data Visualization and Interpretation}
\label{sub:data-interpretation}

Effective communication of spatial information is a key aspect of \gls{gis}, and data visualization is a powerful tool in achieving this goal \cite{longley2015geographic}. Data science and data visualization techniques are employed to create visually appealing and informative representations, allowing the transformation of large and complex spatial datasets into comprehensible and insightful depictions \cite{ki2018gis}.

Well-designed maps and graphics enhance the interpretability of spatial information \cite{slocum2022thematic}, enabling stakeholders to make more informed decisions. \gls{gis} enables the creation of a plethora of spatial data representations---whether in 2D, 3D, or even VR---facilitating the analysis and conveyance of complex spatial patterns and trends in an accessible manner \cite{mitchell1999esri,mitchel2005esri}.

\begin{figure}[h]
  \centering
  \begin{subfigure}{0.45\linewidth}
    \includegraphics[width=\linewidth]{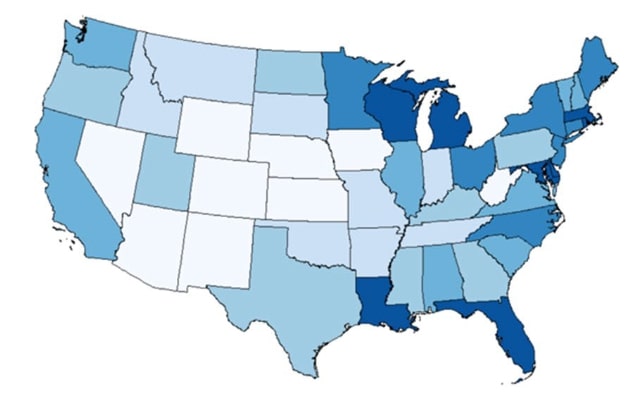}
    \caption{Choropleth Map, adapted from \cite{schiewe2019empirical}.}
    \label{subfig:choropleth}
  \end{subfigure}
  \begin{subfigure}{0.45\linewidth}
    \includegraphics[width=\linewidth]{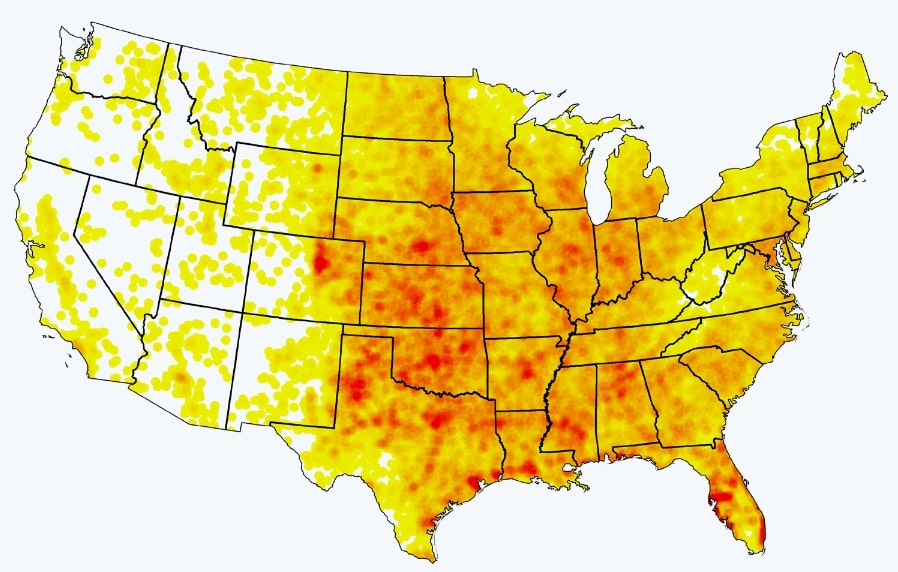}
    \caption{Heat Map, adapted from \cite{deboer2015understanding}}
    \label{subfig:heat}
  \end{subfigure}

  \begin{subfigure}{0.45\linewidth}
    \includegraphics[width=\linewidth]{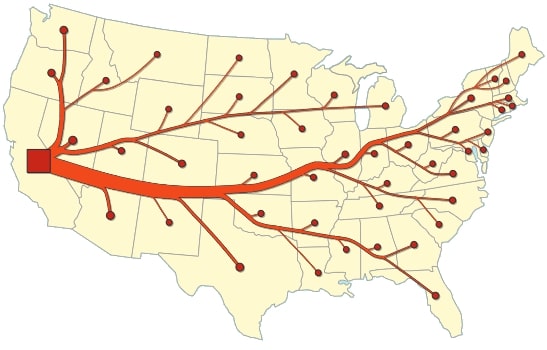}
    \caption{Flow Map, adapted from \cite{buchin2011flow}.}
    \label{subfig:flow}
  \end{subfigure}
  \begin{subfigure}{0.45\linewidth}
    \includegraphics[width=\linewidth]{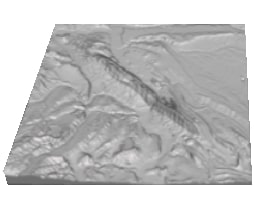}
    \caption{3D Height Model, adapted from \cite{harding2021touchterrain}}
    \label{subfig:terrain}
  \end{subfigure}

  \caption{Examples of \gls{gis} data representation, centered around the continental United States of America. (a) Choropleth maps: thematic maps with areas shaded in proportion to variable values; (b) Heat Maps: represent intensity or concentration of data with colors; (c) Flow maps: visualize movement between locations using lines or arrows; (d) 3D models: represent surfaces in three dimensions, including man-made structures and topographic features.\label{fig:data-representation}}
\end{figure}

Numerous methods exist for representing spatial data. Some of the most important are illustrated in Figure \ref{fig:data-representation}.

\section{Applications}
\label{sec:applications}

The integration of \gls{gis} and data science has spurred transformative applications across various domains. This section will explore some key areas where the synergy of \gls{gis} and data science has been notable, highlighting the versatility and impact of these fields. We start by looking into urban and infrastructure planning (Subsection~\ref{sub:urban-planning}) and examining the far-reaching implications on environmental analysis and conservation efforts (Subsection~\ref{sub:environmental}). Moving forward, we explore the impact on public health and safety (Subsection~\ref{sub:health-safety}). Finally, we discuss several applications within the agricultural domain (Subsection~\ref{sub:agricultural}). This exploration will underscore the multifaceted contributions of \gls{gis} and data science in addressing complex challenges across various sectors.

\subsection{Urban and Infrastructure Planning}
\label{sub:urban-planning}

By analyzing large datasets, urban planners can gain valuable insights into population trends, traffic patterns \cite{yang2019application}, and public infrastructure usage \cite{mohammadi2018enabling}. This information aids in designing efficient transportation systems \cite{torre2018big} and optimizing current city layouts and future developments \cite{hao2015rise}. Furthermore, machine learning algorithms can be employed to predict urban growth, enabling proactive planning to accommodate increasing population demands \cite{chaturvedi2021machine}. Additionally, \gls{gis}-based data science has the potential to help in assessing the impact of urban projects on the environment and social dynamics \cite{kitchin2016ethics}.

\subsection{Environmental Analysis and Conservation}
\label{sub:environmental}

Remote sensing data, satellite imagery, and ground-based sensor data can be integrated to monitor dynamic changes in ecosystems \cite{hansen2012review}, enabling the identification of deforestation and unexpected alterations in habitats \cite{pettorelli2014satellite, sexton2013global}. Furthermore, advanced analytics facilitate the discernment of environmental patterns, allowing conservationists to make well-informed decisions regarding conservation efforts \cite{horning2010remote}. Finally, using historical data, machine learning models can help predict the occurrence and impact of natural disasters \cite{linardos2022machine}, including wildfires \cite{jain2020review, oliveira2023multispectral, fachada-2022}, earthquakes \cite{galkina2019machine}, or floods \cite{mosavi2018flood}. This predictive capability enhances preparedness and response strategies for more effective disaster management.

\subsection{Public Health and Safety}
\label{sub:health-safety}

In the domain of public health, integrated demographic information, disease prevalence, and environmental factors data can be used to identify potential health risks and patterns \cite{kubben2019fundamentals}. Maps can be employed to visualize the spread of diseases, plan healthcare infrastructure, and optimize resource allocation during health emergencies \cite{kamel2020geographical}. Predictive modeling aids in forecasting disease outbreaks, enabling authorities to implement efficient interventions \cite{pullan2011spatial}. Additionally, spatial data science can contribute to identify environmental factors affecting public health, supporting initiatives for cleaner and healthier living environments \cite{maantay2011geospatial}.

\subsection{Agricultural Optimization}
\label{sub:agricultural}

Data science applications in \gls{gis} can significantly enhance agricultural practices by providing insights into crop health, soil conditions, and weather patterns \cite{wardlow2007analysis, mitran2021geospatial,salvado_2019}. Monitoring crop growth, detecting diseases, and optimizing irrigation become feasible through the integration of satellite imagery and terrain sensor data, coupled with predictive algorithms \cite{thenkabail2009global,mestre_2022,classification_2019}. Furthermore, these applications can predict crop yields, identify optimal planting times, determine suitable crop types, and recommend maintenance methods \cite{he2017genotype, priya2001national, dos2023rapid}.

\section{Case Study}
\label{sec:case-study}

In this section, we will present a case study integrating data science and \gls{gis}, describing the data processing and analysis workflow, while highlighting the transformation of raw data into knowledge and insight---from dataset exploration to the final results.

\begin{figure}[h]
\begin{center}
\footnotesize
\begin{tabular}{ccc}
    \centering
    \begin{subfigure}{0.3\linewidth}
      \includegraphics[width=\linewidth]{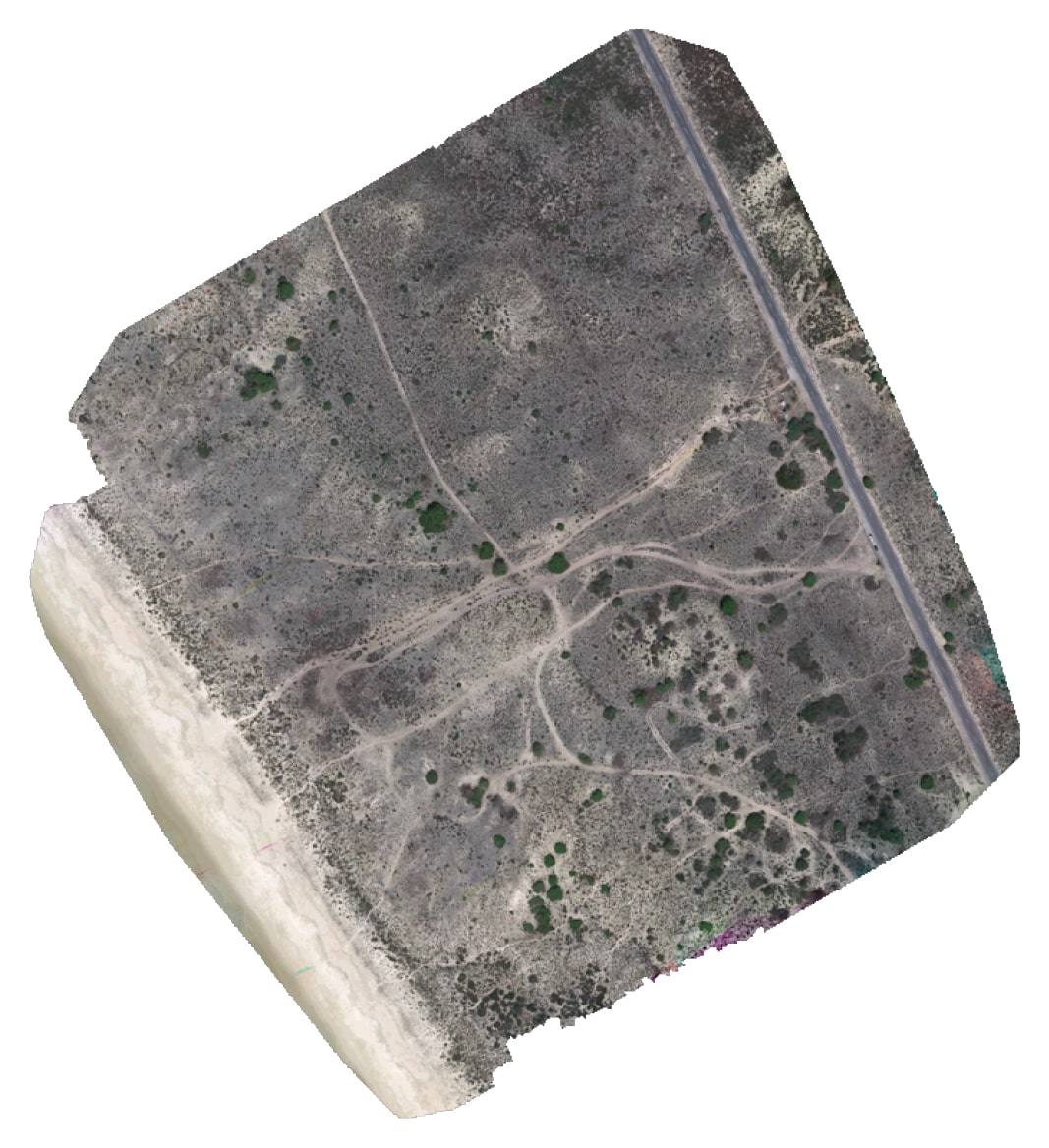}
      \caption{}
      \label{subfig:rgb}
    \end{subfigure} &
    \begin{subfigure}{0.3\linewidth}
      \includegraphics[width=\linewidth]{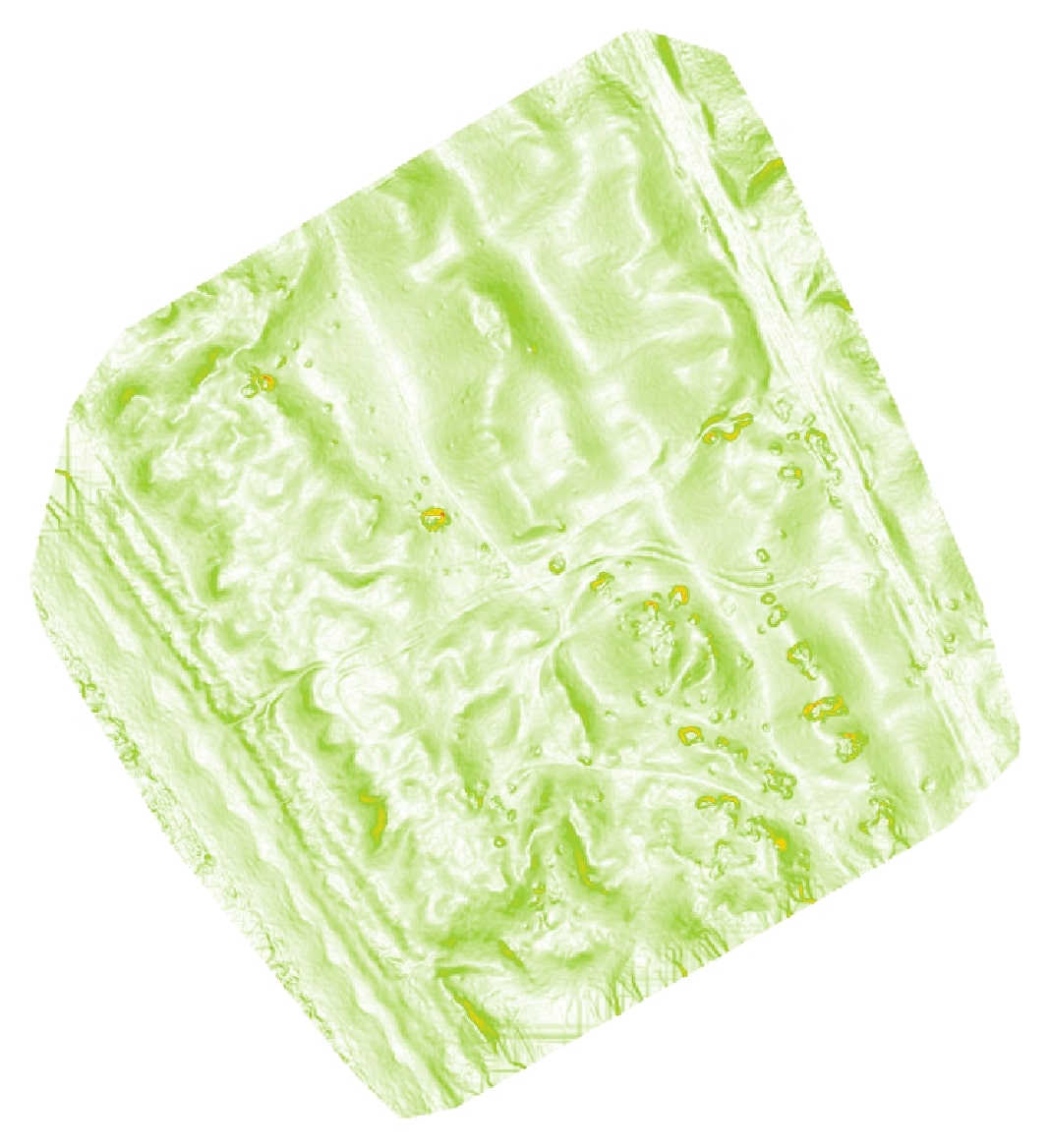}
      \caption{}
      \label{subfig:slope}
    \end{subfigure} &
    \begin{subfigure}{0.3\linewidth}
      \includegraphics[width=\linewidth]{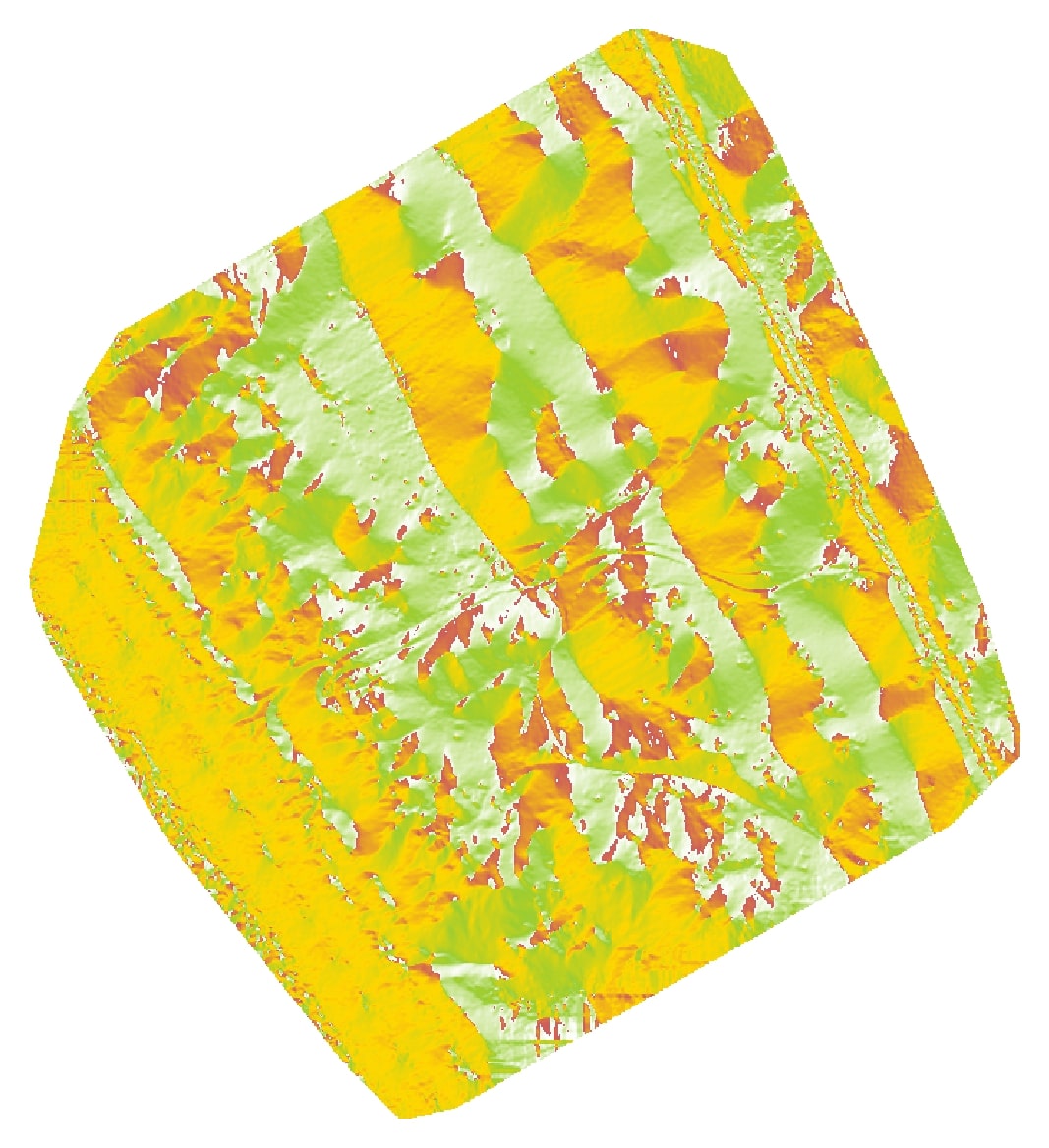}
      \caption{}
      \label{subfig:aspect}
    \end{subfigure} \\

    \begin{subfigure}{0.3\linewidth}
      \includegraphics[width=\linewidth]{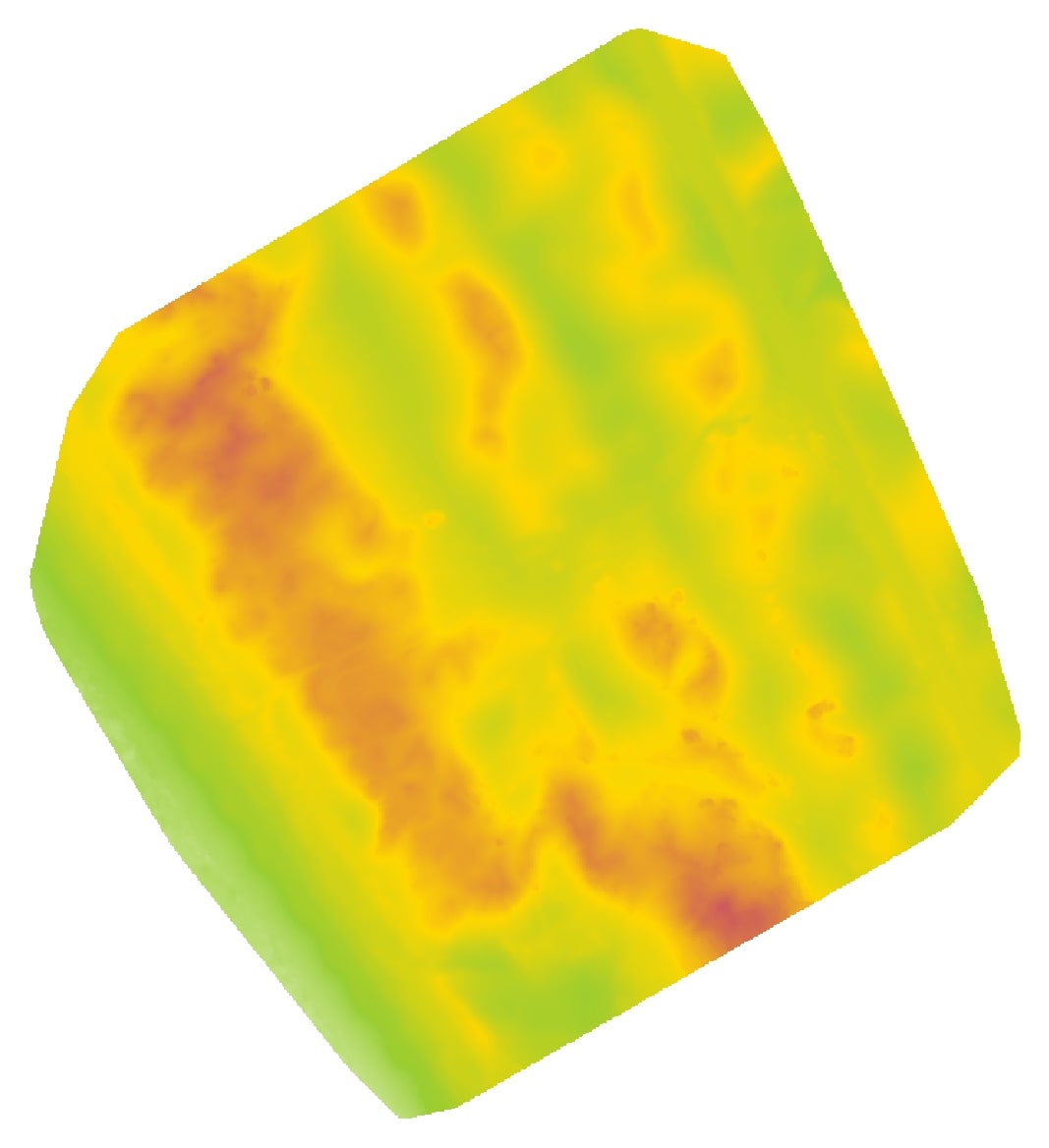}
      \caption{}
      \label{subfig:elevation}
    \end{subfigure} &
    \begin{subfigure}{0.3\linewidth}
      \includegraphics[width=\linewidth]{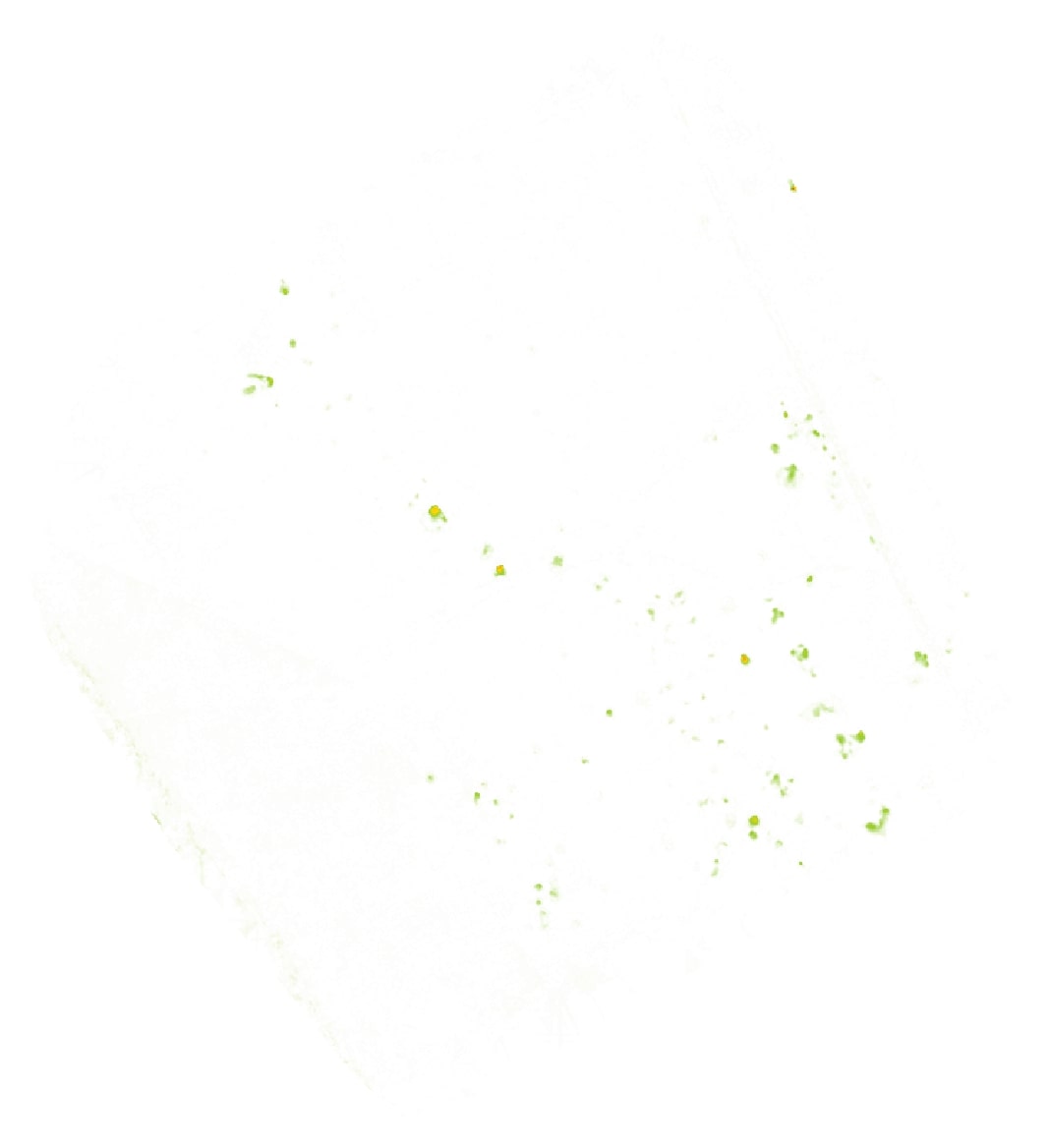}
      \caption{}
      \label{subfig:height}
    \end{subfigure} &
    \begin{subfigure}{0.3\linewidth}
      \includegraphics[width=\linewidth]{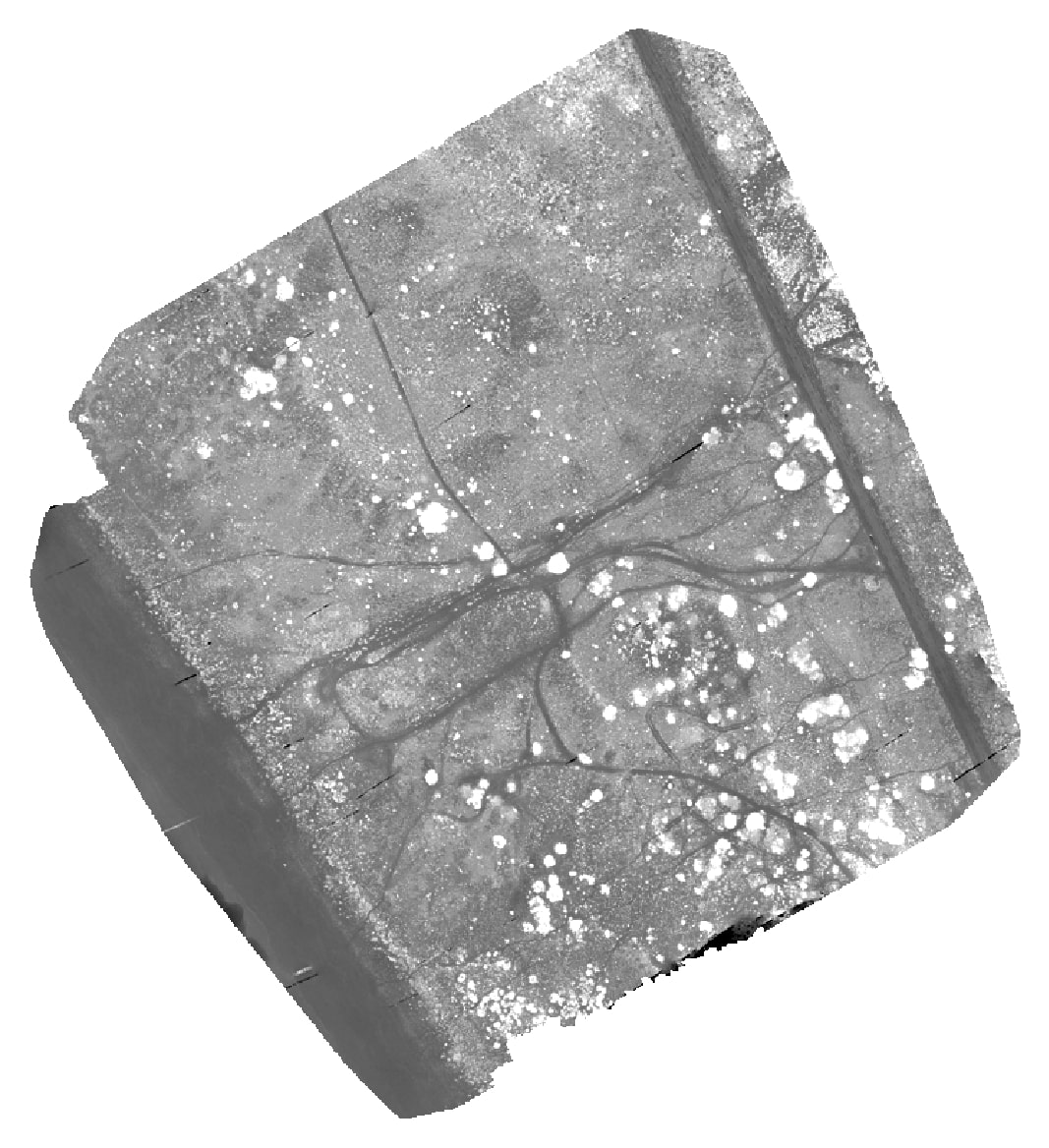}
      \caption{}
      \label{subfig:ndvi}
    \end{subfigure} \\

    \begin{subfigure}{0.3\linewidth}
      \includegraphics[width=\linewidth]{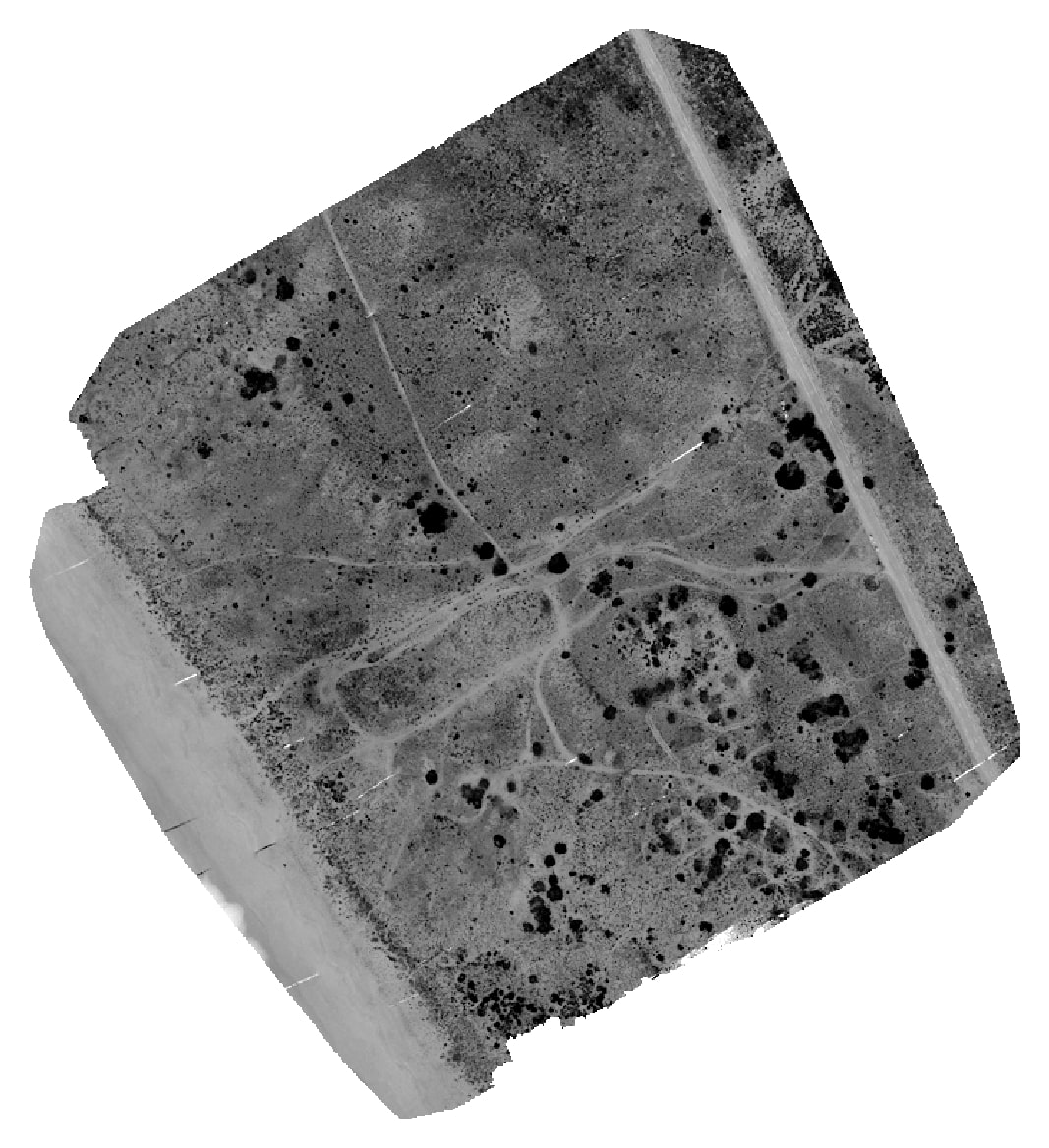}
      \caption{}
      \label{subfig:ndwi}
    \end{subfigure} &
    \begin{subfigure}{0.3\linewidth}
      \includegraphics[width=\linewidth]{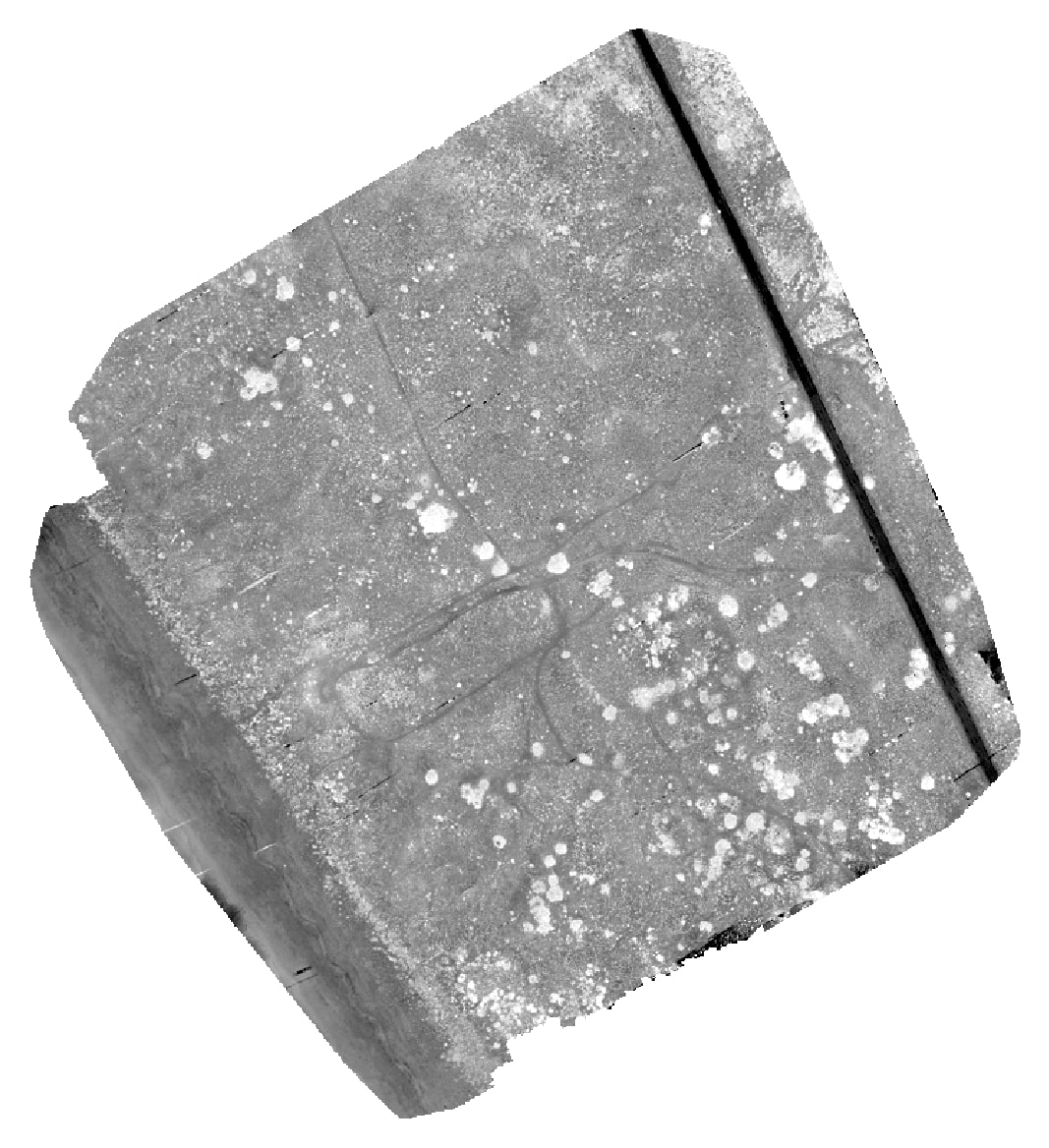}
      \caption{}
      \label{subfig:gari}
    \end{subfigure} &
    \begin{subfigure}{0.3\linewidth}
      \includegraphics[width=\linewidth]{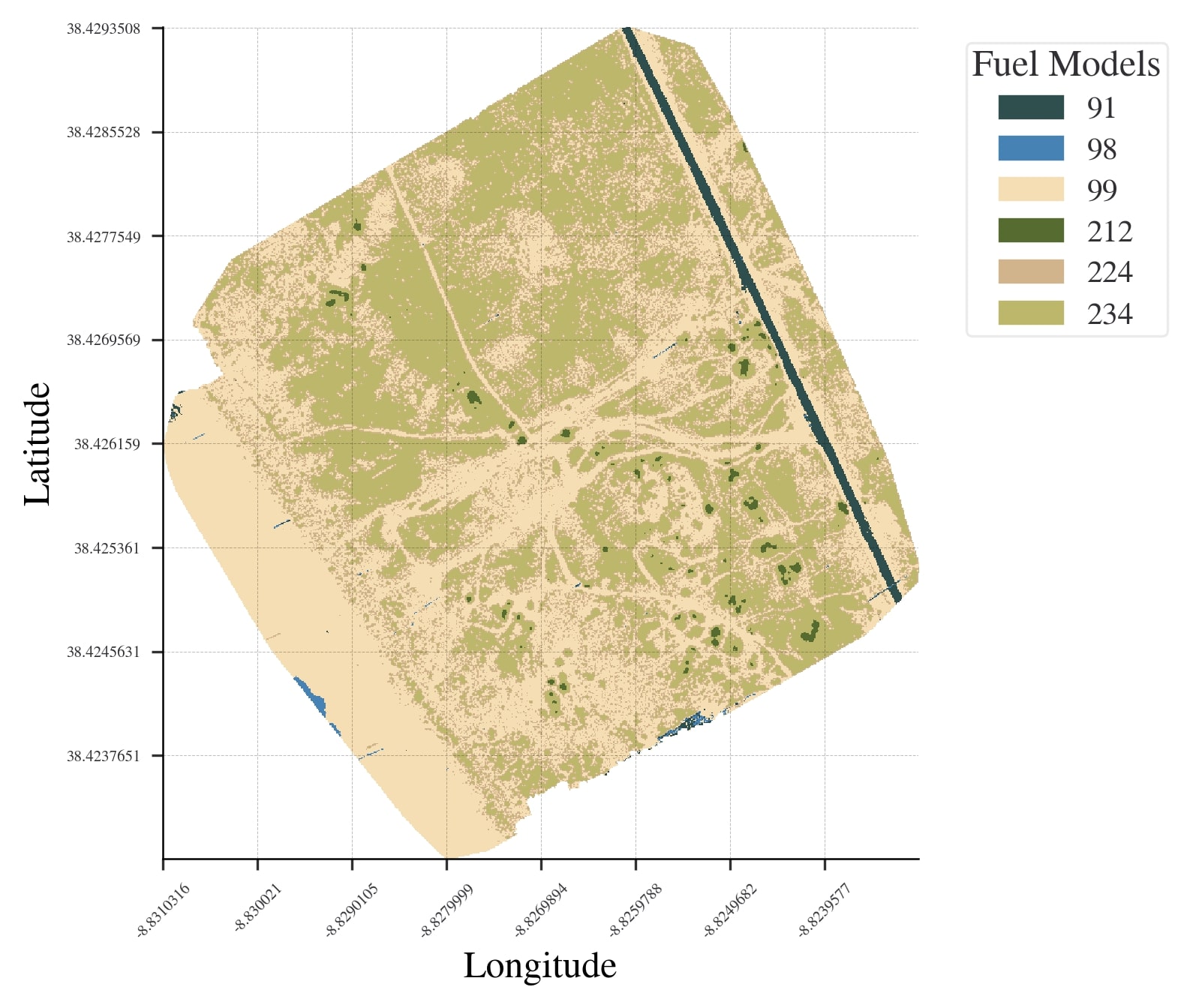}
      \caption{}
      \label{subfig:fuel}
    \end{subfigure} \\

    \begin{subfigure}{0.3\linewidth}
      \includegraphics[width=\linewidth]{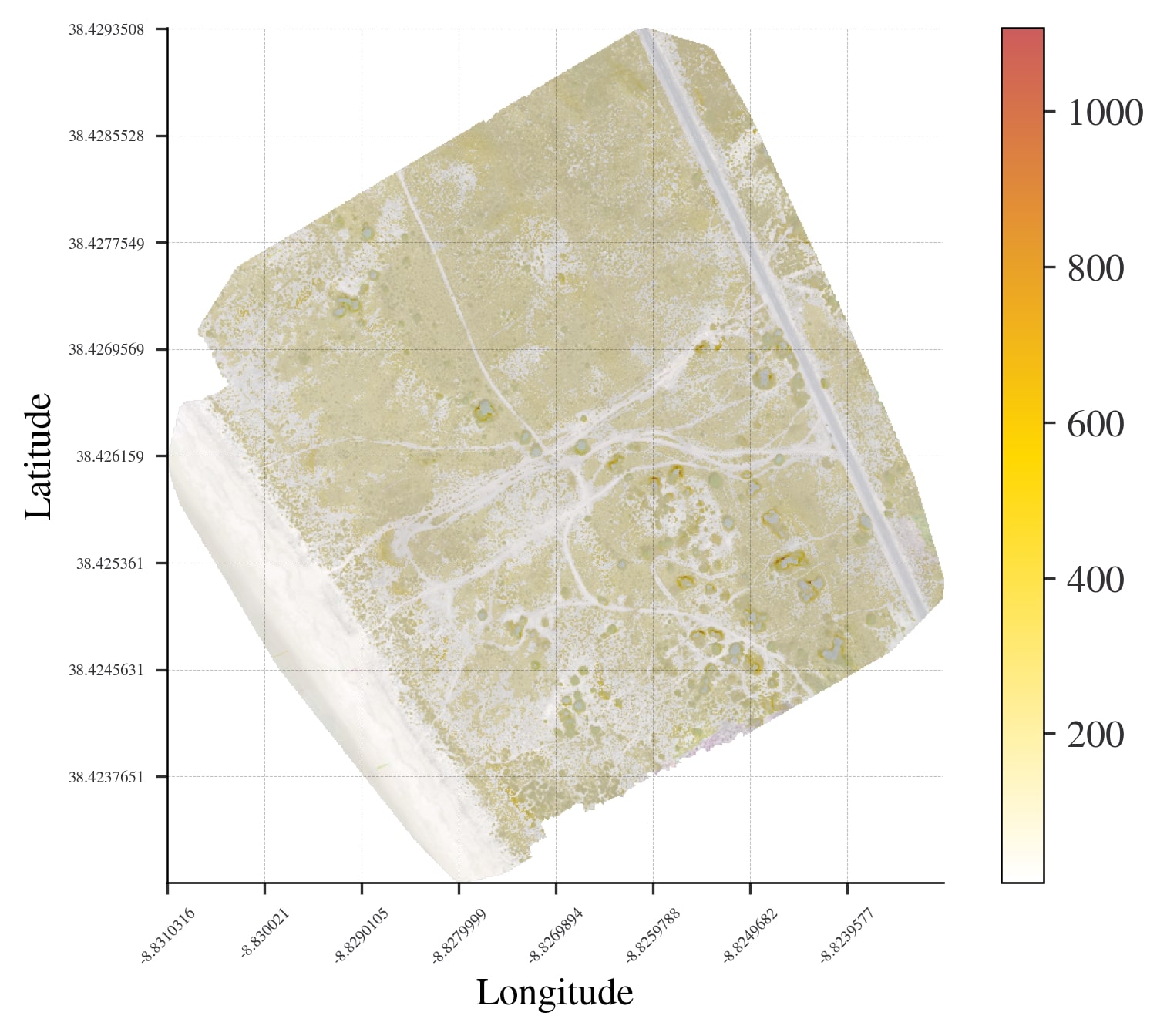}
      \caption{}
      \label{subfig:bi}
    \end{subfigure} & & \\
    \end{tabular}
\caption{Case study results, with a scale of 1 pixel per square meter: a) RGB representation of area of study, in Tróia, Portugal; b) slope in degrees; c) aspect (orientation of slope) in degrees; d) elevation in meters above sea level; e) canopy height in meters; f) vegetation enhanced; g) water features enhanced, with attention drawn to a subtle disturbance at the lower left corner, near the coastline; h) man-made structures; i) fuel map with fuel codes corresponding to a model developed based on \cite{fernandes2021modelos}; and, j) burning Index map generated through simulations using an additional tool and historical weather data. The multispectral data was obtained from~\cite{data_ms_2021,vong_thermal_2022}.\label{fig:case-study}}
\end{center}
\end{figure}

For the purpose of this case study, we use two datasets: 1) aerial imagery encompassing red, blue, green, near-infrared (NIR), and red-edge bands; and, 2) a \gls{dem}. The focus of this study is a small land patch situated in Tróia, Portugal, an RGB representation of which is shown in Figure~\ref{fig:case-study}a. All data was collected in 2022~\cite{precision_farming_2020,fau_2020}.

The initial step involves extracting topographical data from the area using the \gls{dem}. This crucial step yields both the \gls{dtm} and the \gls{dsm}, providing elevation information for each point at ground level and the at maximum height. Some of the most important topographical features to extract are the slope, the aspect, and the elevation of the area. Additionally, the \gls{dem}  enables us to determine vegetation height (as well as the height of other objects) by calculating the difference between the \gls{dtm} and the \gls{dsm}. Figures~\ref{fig:case-study}b,~\ref{fig:case-study}c,~\ref{fig:case-study}d~and~\ref{fig:case-study}e exhibit the data extracted from the DEM dataset.

Subsequently, leveraging the multispectral aerial dataset, we can enhance various objects through the application of multispectral indices. Specifically, we employ the \gls{ndvi} for vegetation, the \gls{ndwi} for water features, and the \gls{gari} for man-made structures, as shown in Figures~\ref{fig:case-study}f,~\ref{fig:case-study}g~and~\ref{fig:case-study}h. These indices were selected empirically, tailored specifically for the nuances of this case study.

Finally, synthesizing all this knowledge, we can create a fuel map of the area, which, in turn, can serve diverse purposes, such as wildfire prevention measures through simulated prediction \cite{fachada-2022}. This could, for example, allow to the computation of a Burning Index (BI) map, serving as a danger indicator to assist authorities in making better decisions regarding wildfire combat and terrain maintenance. Figures~\ref{fig:case-study}i~and~\ref{fig:case-study}j highlight the fuel and BI result for the area analyzed in this case study.

\section{Conclusions}
\label{sec:conclusions}

The integration of data science and \gls{gis} has ushered in a new era of spatial analysis, offering unprecedented opportunities for understanding and interpreting spatial data.

This work has provided an overview of the evolution of both data science and \gls{gis}, tracing their development from their early beginnings to their current state. Furthermore, we have highlighted the key components defining the intersection of these fields and have explored a range of applications demonstrating the versatility and impact of \gls{gis} and data science integration.

The presented case study exemplifies the practical implementation of data science techniques in \gls{gis}, showcasing how raw data can be transformed into actionable insights. Furthermore, we illustrated the application of this knowledge in creating fuel maps and predicting burning indices for wildfire prevention measures.

Looking ahead, the prospects for further research and development in this field are promising. As technologies continue to advance, there is an opportunity to explore new methodologies and algorithms for enhanced spatial analysis and decision support systems. Furthermore, the integration of emerging technologies such as artificial intelligence, remote sensing, and \gls{iot} devices holds immense potential for addressing pressing global challenges, including climate change, natural disasters, and sustainable development.

\bibliographystyle{unsrt}

\end{document}